\documentclass[12pt]{iopart}
\usepackage{amssymb}
\usepackage{setstack}
\usepackage{verbatim}
\usepackage{bm}
\usepackage{graphicx}

\begin{document}

\newcommand{\ket}[1]{| #1 \rangle}
\newcommand{\bra}[1]{\langle #1 |}
\newcommand{\smallfrac}[2]{\mbox{$\frac{#1}{#2}$}}
\newcommand{\arccosh}{\mathrm{arccosh}\,}

\title[The characteristic polynomial of the next-nearest-neighbour qubit chain]{The characteristic polynomial of the next-nearest-neighbour qubit chain for single excitations}
\author{C J Mewton and Z Ficek}
\eads{\mailto{mewton@physics.uq.edu.au}, \mailto{ficek@physics.uq.edu.au}}
\address{Department of Physics, School of Physical Sciences,
The University of Queensland, Brisbane, Queensland 4072, Australia}
\date{\today}

\begin{abstract}
The characteristic polynomial for a chain of dipole-dipole coupled two-level atoms with nearest-neighbour and next-nearest-neighbour interactions is developed for the study of eigenvalues and eigenvectors for single-photon excitations.  Such a system is mathematically equivalent to an $XX$ spin chain in an external magnetic field.  We find the exact form of the polynomial in terms of the Chebyshev polynomials of the second kind that is valid for an arbitrary number of atoms and coupling strengths.  We then propose a technique for expressing the roots of the polynomial as a power series in the coupling constants.  The general properties of the solutions are also explored, to shed some light on the general properties that the exact, analytic form of the energy eigenvalues should have.  A method for deriving the eigenvectors of the Hamiltonian is also outlined.
\end{abstract}



\section{Introduction}

The properties and dynamics of linear chains of atoms under the nearest-neighbour coupling approximation have been the subject of considerable interest for many 
decades~\cite{bib:Lieb1961,bib:Katsura1962,bib:Katsura1963}. Recently, a new interest has arisen 
in linear chains in the context of quantum information and computation~\cite{bib:Bose2007}, where linear chains have been proposed as entanglement carriers that can be relatively easily realized in practice by diverse physical systems such as cold atoms trapped in optical potentials~\cite{bib:Dorner2003, bib:Garcia-Ripoll2004}, quantum dots~\cite{bib:Loss1998, bib:Burkard1999, bib:Imamoglu1999} and Josephson junction arrays~\cite{bib:Makhlin2001}.
Despite the simplification that takes into account only the nearest neighbour interaction between the atoms, the analysis of the properties of a linear chain containing a large number of atoms are still complex. Therefore most of the developments have been focused on the analysis of the energy and entanglement of the ground state only. 
Numerical results for the ground state energy and two-particle correlation functions were obtained using the modified Lanczos method~\cite{bib:Mancini1994, bib:Massano1992}. Some approximate analytic results for the ground state energy and higher excitations were also reported~\cite{bib:Gottlieb1991}.   In addition, the thermodynamic properties of a frustrated Heisenberg chain were studied~\cite{bib:Gluzman1994} and fluctuations in the axial Ising model were analysed~\cite{bib:Harris1995}.
More recently, numerical investigations were conducted~\cite{bib:Otsuka2002}, and also the scaling of excitations were analysed~\cite{bib:Controzzi2005, bib:Kumar2007}.  The work of Lamas {\it et al.}~\cite{bib:Lamas2006} have introduced approximations for handling a spin chain with different coupling strengths along the chain. 
Apart from numerical results, an analytic approach of finding the ground state was proposed \cite{bib:Kohler1997}, and the properties of frustrated Heisenberg chains were also studied \cite{bib:Ivanov1998}.  

As we have already mentioned, exact treatments of linear or closed chains are limited to the studies of energy and entanglement of the ground state or to a small number of atoms of an arbitrary 
excitation~\cite{bib:Freedhoff2004, bib:Rudolph2004}. In the present work, we consider the dipole-dipole chain, which is equivalent to an $XX$ chain in a magnetic field, and derive the characteristic polynomial equation for the energy states and investigate the properties of its solutions. Derivation of the eigenstates of a chain of interacting atoms is of great importance for the study of entanglement creation, its transfer through the chain and stability against the decoherence. We also go beyond the nearest-neighbour coupling approximation to include the next nearest-neighbour coupling. 

It is generally believed that the nearest-neighbour approximation is a correct approach to study the properties and dynamics of linear chains. However, for realistic physical implementations of linear chains some directions of investigations are important such as the role of long range dipole-dipole interactions. It is the main purpose of this paper to address this issue by considering effects of the next-nearest-neighbour interactions. We shall deal with the eigenvalue equation for a chain of $n$ coupled atoms and extract the characteristic polynomial for the eigenvalue equation of the Hamiltonian of the system for a single excitation.  We then propose a technique of expressing the roots of the polynomial in terms of a power series in the coupling constants.
Deriving the exact characteristic polynomial moves the research field closer to the goal of exact, analytic results for a chain of arbitrary length and arbitrary excitation, for it is the roots of this polynomial which give the energy eigenvalues of the chain.

\section{System Hamiltonian}

\begin{figure}
\begin{center}
\includegraphics{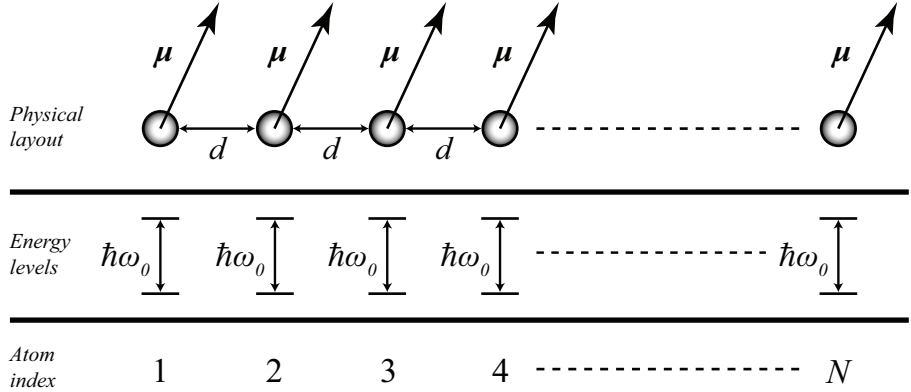}
\end{center}
\caption{Physical layout of the linear chain.  The atoms are equally separated from each other by a distance $d$, and are polarized in the same direction in their dipole moments $\boldsymbol{\mu}$ and their two-level energy structure (shown beneath each atom).}
\label{fig:linearchain_schematic}
\end{figure}

We consider a linear chain of $n$ identical two-level atoms.  The atoms are assumed to be equally spaced and confined to fixed positions, and each atom can only interact with its nearest neighbours 
through the dipole-dipole interaction.  A schematic of the system is shown in Figure \ref{fig:linearchain_schematic}.

The dynamics of the system are governed by the Hamiltonian:
\begin{equation}
H = \sum_{i=1}^{n} \omega_0 S_z^i + a \sum_{|i-j|=1} S^+_i S^-_j+ b \sum_{|i-j|=2} S^+_i S^-_j,
\label{eq:bnn_h}
\end{equation}
where the parameters $a$ and $b$ describe the nearest-neighbour and next-nearest-neighbour coupling constants, respectively.  The operators $S_i^+$ and $S_i^-$ raise and lower the state of the $i$-th atom by a single excitation, and $S_i^z = S_i^+S_i^-$.  This definition of $S_i^z$ makes the energy of the ground state of the system equal to zero.  We first briefly show that the above Hamiltonian is equivalent to an $XX$ chain in a magnetic field \cite{bib:Lieb1961, bib:Katsura1962}.  We first express the raising and lowering operators in terms of Pauli matrices:
\begin{equation}
S_k^+ = \smallfrac{1}{2} (\sigma_k^x + i \sigma_k^y), \quad S_k^- = \smallfrac{1}{2}(\sigma_k^x - i \sigma_k^y).
\end{equation}
These relations identify the internal states of a two-level atom with the $z$ component of spin for a spin-$\frac{1}{2}$ particle.  Substituting these relations into (\ref{eq:bnn_h}) and making use of the commutation relations for Pauli matrices gives the analogous Hamiltonian for a spin chain:
\begin{equation}
\fl H = \frac{1}{2} n \omega_0 + \frac{1}{2} \omega_0 \sum_{i=1}^{n} \sigma_i^z + \frac{1}{4} a \sum_{|i-j|=1} ( \sigma_i^x \sigma_j^x + \sigma_i^y \sigma_j^y)
+ \frac{1}{4} b \sum_{|i-j|=2} (\sigma_i^x \sigma_j^x + \sigma_i^y \sigma_j^y).
\end{equation}
Ignoring the first term, which just shifts the energy scale, the second term represents the coupling of an external magnetic field to the spins of the particles, while the second and third bracketed terms represent $XX$ coupling for nearest and next-nearest neighbours.

It is important to note that our work does not depend on specific features of the dipole-dipole interaction; all of our results are therefore also applicable to spin chains.  We will, however, give specific examples in terms of the dipole-dipole coupled chain to fix our ideas, since it provides an easy way to think of a system described by couplings between raising and lowering operators.

\begin{figure}
\begin{center}
\includegraphics{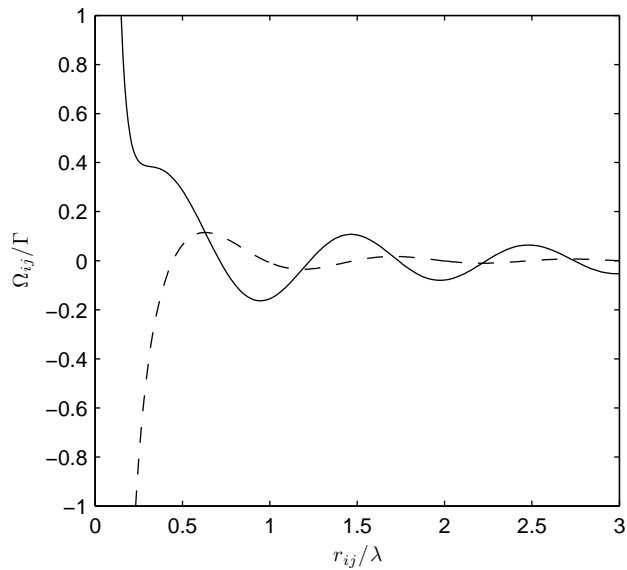}
\end{center}
\caption{Dipole-dipole interaction strength as a function of $r_{ij}/ \lambda_{0}$ for two different orientations of the atomic dipole 
moments relative to the interatomic axis, $\hat{\boldsymbol{\mu}} \bot \hat{\boldsymbol{r}}_{ij}$ (solid line) and 
$\hat{\boldsymbol{\mu}} \parallel \hat{\boldsymbol{r}}_{ij}$ (dashed line).}
\label{fig:dipolegraph}
\end{figure}

Although we will regard $a$ and $b$ as arbitrary parameters, their magnitudes can be explicitly determined, for the case of dipole-dipole coupling, from the expression~\cite{bib:Lehmberg1970, bib:Ficek2002}
\begin{equation}
\fl \Omega_{ij}  =  \frac{3}{4}  \Gamma \left\{
 -[ 1 - (\hat{\boldsymbol{\mu}} \cdot \hat{\boldsymbol{r}}_{ij})^2 ] \frac{\cos x_{ij}}{x_{ij}} 
 + [1 - 3(\hat{\boldsymbol{\mu}} \cdot \hat{\boldsymbol{r}}_{ij})^2]
 \left[ \frac{\sin x_{ij}}{x_{ij}^2} + \frac{\cos x_{ij}}{x_{ij}^3} \right] \right\} ,
\end{equation}
where $x_{ij} = 2\pi r_{ij}/\lambda_{0}$ is the distance between the atoms $i$ and $j$ relative to the resonant wavelength $\lambda_{0} = c/2 \pi \omega_0$ of the atomic transition, $\Gamma$ is the decay rate of the atomic transition, assumed the same for all atoms, and  $\hat{\boldsymbol{\mu}}$ and  $\hat{\boldsymbol{r}}_{ij}$ are unit vectors in the direction of the atomic dipole moments and the displacement $\boldsymbol{r}_{ij}$ between the $i$-th and $j$-th atoms.
The dependence of $\Omega_{ij}$ on the distance between two arbitrary atoms in the chain is illustrated in 
Figure \ref{fig:dipolegraph}.  In the present case, $a = \Omega_{i,i+1}$ and $b = \Omega_{i, i+2}$.

We will first derive the eigenvalue equation for the Hamiltonian and investigate is its symmetry properties.  This will be followed by a discussion on the form of the eigenstates of the Hamiltonian.

\section{Eigenvalue equation}

Since our system comprises of $n$ two-level atoms, it is clear that up to $n$ photons may be present in the chain.  We will consider only one-photon states due to the complexity of the problem for higher-order excitations.  Single-photon states may be useful in approximating higher-order states.

Our goal is to find analytical expressions for the one-photon eigenstates and their energies, which are found by solving the
eigenvalue equation $H\ket{\psi} = E \ket{\psi}$, where $H$ is given in (\ref{eq:bnn_h}). We proceed by writing the 
characteristic equation
\begin{equation}
T_n \equiv \det(H -E I) = \left| \begin{array}{cccccc}
\lambda & a & b & \phantom{\ddots} & \phantom{\ddots} & \phantom{\ddots} \\
a & \lambda & a & b & \phantom{\ddots} & \\
b & a & \lambda & a & \ddots & \\
\phantom{\ddots} & b & a & \lambda & \ddots & b\\
\phantom{\ddots} & \phantom{\ddots} & \ddots & \ddots & \ddots & a \\
\phantom{\ddots} & \phantom{\ddots} & \phantom{\ddots} & b & a & \lambda
\end{array} \right|_{n},
\label{eq:TnDef}
\end{equation}
where $\lambda = \omega_0 - E$ and $I$ is the $n \times n$ identity matrix.  The matrix $T_{n}$, which represents the 
eigenvalue equation, is a $n$-dimensional five-diagonal matrix with two off-diagonals set by the nearest-neighbour coupling and the other two off-diagonal set by the next-nearest-neighbour coupling.
For clarity, all entries in a determinant which are zero are simply omitted, and the order of the determinant (in this case, $n$) is placed as a subscript on the bottom right-hand corner of the determinant.

The eigenvalues $E$ are then given by the roots of $T_n$, i.e. those values of $E$ which make $T_n(E) = 0$.  At this point, we could pick a specific numerical value for $n$ and solve for the roots of $T_n$ by expanding out the determinant.  However, we would like to keep our results as general as possible; we therefore seek an alternate expression for $T_n$ such that it is an analytic function of $n$.  Such a form has the advantage that, as evidenced by the final result, one does not need to construct a matrix to find the eigenvalues.  This can be an advantage particularly when $n$ is large since less computer memory is required for the calculation.

To find a form of $T_n$ that is analytic in $n$, we first expand the determinant appearing in (\ref{eq:TnDef}) to find a recurrence relation for $T_n$.  An expansion of the determinant leads to a closed recurrence equation
\begin{equation}
T_n = \lambda T_{n-1} - a D_{n-1} + ba D_{n-2} - b^2 \lambda T_{n-3} + b^4 T_{n-4} ,
\label{eq:TnDnRecurrence1}
\end{equation}
where
\begin{eqnarray}
D_n & \equiv & \left| \begin{array}{cccccc}
a & a & b & \phantom{\ddots} & \\
b & \lambda & a & b & \phantom{\ddots}\\
\phantom{\ddots} & a & \lambda & a & \ddots\\
\phantom{\ddots} & b & a & \lambda & \ddots & b\\
\phantom{\ddots} & \phantom{\ddots} & \ddots & \ddots & \ddots & a\\
\phantom{\ddots} & \phantom{\ddots} & \phantom{\ddots} & b & a & \lambda
\end{array} \right|_n .
\end{eqnarray}

The next step is to obtain an expression for $T_n$ that does not have any $D_{n}$ terms in it.  We do this by finding a recurrence relation for $D_n$.  Having then two simultaneous recurrence relations involving $T$'s and $D$'s, we can then eliminate all the $D$'s in the recurrence equation (\ref{eq:TnDnRecurrence1}) leading to a recurrence relation solely in terms of $T$'s.

Thus, expanding the determinant $D_n$ along the first column gives
\begin{equation}
D_n = a T_{n-1} - b D_{n-1}.
\label{eq:TnDnRecurrence2}
\end{equation}
It is seen why we insisted on not expanding out the $D$'s terms in (\ref{eq:TnDnRecurrence1}). This procedure would be equivalent to repeatedly substituting (\ref{eq:TnDnRecurrence2}), which would not end until all the rows of $T_n$ have been Laplace expanded.  This would give us an effectively useless $n$-term recurrence relation since the number of terms grows with $n$.

We now use (\ref{eq:TnDnRecurrence2}) to eliminate the $D_{n-1}$ term from (\ref{eq:TnDnRecurrence1}).  
This step gives
\begin{equation}
T_n = \lambda T_{n-1} + 2ba D_{n-2} - a^2 T_{n-2} - b^2 \lambda T_{n-3} + b^4 T_{n-4}.
\end{equation}
Next, by adding $bT_{n-1}$ to both sides of the above equation, we find
\begin{eqnarray}
T_n + b T_{n-1} & = & \lambda ( T_{n-1} + b T_{n-2} ) + 2ba (D_{n-2} + b D_{n-3} ) \nonumber \\
&&  - a^2 (T_{n-2} +  b T_{n-3} ) - b^2 \lambda (T_{n-3} + b T_{n-4} ) \nonumber \\
&& + b^4 (T_{n-4} + b T_{n-5} ) . \label{eq:TnAlmostThere}
\end{eqnarray}
Since $D_{n-2} + b D_{n-3} = a T_{n-3}$, as it is seen from  (\ref{eq:TnDnRecurrence2}), we finally arrive 
to the recurrence relation for $T_{n}$:
\begin{eqnarray}
T_n & = & (\lambda - b) T_{n-1} + (\lambda b - a^2) T_{n-2} + (a^2b - b^2 \lambda) T_{n-3} \nonumber \\
&& + (b^4 - b^3 \lambda) T_{n-4} + b^5 T_{n-5}.
\label{eq:TnRecurrence}
\end{eqnarray}
The recurrence relation (\ref{eq:TnRecurrence}) looks complicated and does not reveal any similarities to the well known 
recurrence relations for special functions. Therefore, we will try to find a non-recursive form for $T_n$.

\section{Characteristic equation for $T_n$}

Since the recurrence relation (\ref{eq:TnRecurrence}) is a linear homogeneous equation with constant coefficients, 
we let $T_n = x^n$ to form the characteristic equation
\begin{eqnarray}
x^5 - (\lambda - b)x^4 - (\lambda b -a^2)x^3 - (a^2b-b^2 \lambda)x^2  -(b^4-b^3 \lambda) x - b^5 = 0 .
\end{eqnarray}
The roots of the characteristic equation can be easily found, and are of the form
\begin{equation}
x = \left\{ \begin{array}{l}
b\\
x_{\pm} \equiv \frac{1}{2}\chi_{+} \pm \frac{1}{2} \sqrt{ \chi_{+}^2 - 4b^2 }\\
y_{\pm} \equiv \frac{1}{2}\chi_{-} \pm \frac{1}{2} \sqrt{ \chi_{-}^2 - 4b^2 }\\
\end{array} \right. ,\label{e13}
\end{equation}
where
\begin{equation}
\chi_{\pm} \equiv \frac{\lambda -2b \pm \Delta}{2}
\end{equation}
and
\begin{equation}
\Delta \equiv \sqrt{(\lambda + 2b)^2 - 4a^2}.
\end{equation}
Thus, with the roots (\ref{e13}), the general solution to the recurrence relation (\ref{eq:TnRecurrence}) can be written as
\begin{equation}
T_n = Gb^n + P_{+}x_{+}^n + P_{-}x_{-}^n + Q_{+}y_{+}^n + Q_{-}y_{-}^n,
\label{eq:TnGeneralArb}
\end{equation}
where the parameters $G$, $P_{\pm}$, and $Q_{\pm}$ are arbitrary constants.  These constants are determined by requiring that (\ref{eq:TnGeneralArb}) gives the correct result for five different values of $n$. We chose $n=1$ through to $5$ inclusive (the specific choices for $n$ does not matter; the constants will not change).  This gives us a set of five simultaneous equations or ``boundary conditions'' in the five unknowns $G$, $P_{\pm}$, and $Q_{\pm}$, which is sufficient to give unique values for these parameters.  These boundary conditions force the five arbitrary constants to take on the following values:
\begin{eqnarray}
G &=& {\frac {-2 {b}^{2}}{4b \lambda-8b^2-a^2}} ,\\
P_{\pm}  &=& \pm \frac{\lambda ( \lambda+2b+\Delta ) -2a^2}{2 \Delta \sqrt{\chi_{+}^2 -4b^2}} + \frac{\chi_{+}+b}{2\Delta} 
 + \frac{b[ 2b( \Delta-2b -3 \chi_{+} )+a^2]}{2 \Delta  (a^2 + 8b^2 -4b \lambda)} ,\label{eq:Ppm}\\
Q_{\pm}  &=&  \mp \frac{\lambda ( \lambda+2b-\Delta ) -2a^2}{2 \Delta \sqrt{\chi_{-}^2 -4b^2}} - \frac{\chi_{-}+b}{2\Delta}
- \frac{b[ 2b( \Delta +2b +3 \chi_{-} )-a^2]}{2 \Delta  (a^2 + 8b^2 -4b \lambda)} .\label{eq:Qpm}
\end{eqnarray}
In principle, we have found the expression for $T_{n}$ that we were looking for, i.e. one that is a function of $n$.  However, it is somewhat unwieldy, and we will see in the next subsection that it can indeed be simplified.

\section{Simplification procedure}

The solution  (\ref{eq:TnGeneralArb}) can be simplified to a compact form.  To show this, let us consider the ordinary power series generating function of $T_n$:
\begin{equation}
\mathfrak{T}(z) \equiv \sum_{n=0}^{\infty} T_n z^n.
\end{equation}
From (\ref{eq:TnGeneralArb}), we clearly have the equality
\begin{equation}
\mathfrak{T}(z) = \frac{G}{1-bz} + \frac{P_{+}}{1-x_{+} z} + \frac{P_{-}}{1-x_{-} z} + \frac{Q_{+}}{1-y_{+} z} + \frac{Q_{-}}{1-y_{-} z},
\end{equation}
which, after the substitution of (\ref{eq:Ppm}) and (\ref{eq:Qpm}), becomes
\begin{equation}
\mathfrak{T}(z) = \frac{1+bz}{(1-bz)(1-x_{+} z)(1-x_{-} z)(1-y_{+} z)(1-y_{-} z)}.
\end{equation}
Here we see the remarkable result that despite the complexity of the coefficients $P_{\pm}$ and $Q_{\pm}$ in (\ref{eq:Ppm}) and (\ref{eq:Qpm}), the resulting generating function is relatively simple.  We may proceed further and find that it may be expressed as follows:
\begin{equation}
\mathfrak{T}(z) = \frac{1+bz}{1-bz} \frac{1}{\Delta z} \left( \frac{1}{1 - \chi_{+} z + b^2 z^2} - \frac{1}{1 - \chi_{-} z + b^2 z^2} \right).
\end{equation}
One can recognize that the two terms enclosed by parentheses are the ordinary power series generating functions of the Chebyshev polynomials of the second kind \cite{bib:Rivlin1974}:
\begin{equation}
U_n( \cos \theta ) \equiv \frac{\sin (n+1) \theta}{\sin \theta}.
\end{equation}
We can therefore write
\begin{equation}
\mathfrak{T}(z) = \frac{1+bz}{1-bz} \frac{1}{\Delta z} \sum_{n=0}^{\infty} \left[ U_n \left( \frac{\chi_{+}}{2b} \right) - U_n \left( \frac{\chi_{-}}{2b} \right) \right] b^n z^n.
\end{equation}
Expanding $1/(1-bz)$ and using the Cauchy product gives
\begin{equation}
\mathfrak{T}(z) = \frac{1+bz}{\Delta z} \sum_{n=0}^{\infty} \left\{ \sum_{k=0}^{n} \left[ U_k \left( \frac{\chi_{+}}{2b} \right) - U_k \left( \frac{\chi_{-}}{2b} \right) \right] \right\} b^n z^n.
\end{equation}
The function $T_n$ may thus be expressed as follows:
\begin{eqnarray}
T_n & = & \frac{b^{n+1}}{\Delta} \left\{\sum_{k=0}^{n+1} \left[ U_k \left( \frac{\chi_{+}}{2b} \right) - U_k \left( \frac{\chi_{-}}{2b} \right) \right] 
+  \sum_{k=0}^{n} \left[ U_k \left( \frac{\chi_{+}}{2b} \right) - U_k \left( \frac{\chi_{-}}{2b} \right) \right]\right\}
\nonumber \\
&=& \frac{b^{n+1}}{\Delta} \left\{ 2 \sum_{k=0}^{n} \left[ U_k \left( \frac{\chi_{+}}{2b} \right) - U_k \left( \frac{\chi_{-}}{2b} \right) \right] \right. \nonumber \\
&&\left. + \left[ U_{n+1} \left( \frac{\chi_{+}}{2b} \right) - U_{n+1} \left( \frac{\chi_{-}}{2b} \right) \right] \right\} .
\end{eqnarray}

Given the form of the Chebyshev polynomials $U_k$ in terms of trigonometric functions, we can perform the summation in the above expression over $k$ using the formula for the summation of a geometric series.  Following this procedure, we get
\begin{equation}
T_n = \frac{b^{n+1}}{\Delta} F_n(\theta, \phi),
\end{equation}
where
\begin{equation}
F_n(\theta, \phi) \equiv \frac{1 - \cos\left[(n+2) \theta\right]}{1-\cos \theta} - \frac{1 - \cos\left[(n+2) \phi\right]}{1 - \cos \phi},
\label{eq:FnThetaPhi}
\end{equation}
and the angles are given by
\begin{equation}
\theta \equiv \arccos \frac{\chi_{+}}{2b}, \quad \phi \equiv \arccos \frac{\chi_{-}}{2b}.
\end{equation}

Using the half-angle formula for the cosine function, the expression for $F_n (\theta, \phi)$ reduces to a form
\begin{equation}
F_n (\theta, \phi) = \left[ U_{n+1} \left(\cos \frac{\theta}{2} \right) \right]^2
-  \left[ U_{n+1} \left(\cos \frac{\phi}{2} \right) \right]^2.
\end{equation}
To make the relationship between $\theta$ and $\phi$ in the above expression more illuminating, let us define the variable $\alpha$ such that
\begin{equation}
\lambda + 2b = 2a \cos 2 \alpha.
\end{equation}
Since the Hamiltonian of the system is Hermitian, its eigenvalues must be real. Hence $\lambda$ is always real.  
This requires $\alpha$ to be of the form $x$, $ix$ or $\pi/2 + ix$, where $x \in \mathbb{R}$.  
In terms of $\alpha$, we find that 
\begin{eqnarray}
\cos \frac{1}{2} \theta = \gamma \rme^{\rmi \alpha} \quad {\rm and}\quad  \cos \frac{1}{2} \phi = \gamma \rme^{-\rmi \alpha} , 
\end{eqnarray}
where 
\begin{equation}
\gamma \equiv \sqrt{\frac{a}{4b}}.
\label{eq:gammadef}
\end{equation}

We thus have arrived at an expression for $T_n$ in a very simple form given in terms of Chebyshev polynomials of the second kind:
\begin{equation}
T_n = \frac{-\rmi b^{n+1}}{2a \sin 2 \alpha} \left\{ \left[ U_{n+1} \left( \gamma \rme^{\rmi \alpha} \right) \right]^2
-  \left[ U_{n+1} \left( \gamma \rme^{-\rmi \alpha} \right) \right]^2 \right\}. \label{eq:TnFinalForm}
\end{equation}
The relation given in (\ref{eq:TnFinalForm}) is the principal result of the paper. Its importance for our work is 
that we have turned the determinantal form of $T_n$ into an expression which is an analytic function of $n$ if we make use of the trigonometric definition of the Chebyshev polynomials.  If the exact roots of this equation could be found, we would have the explicit energy eigenvalues for single-photon states, valid for arbitrary chain lengths and coupling constants. Despite this  complication, we can obtain the explicit energy eigenvalues and eigenvectors for some special cases.

\section{The case of no nearest-neighbour interaction}

\begin{figure}
\begin{center}
\includegraphics{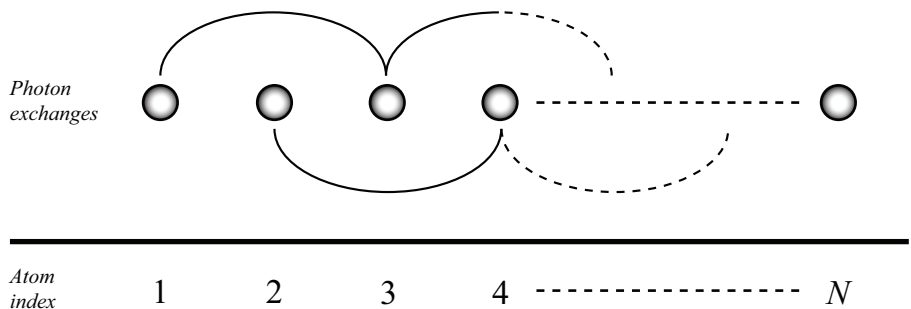}
\end{center}
\caption{When atoms can only interact via next-nearest-neighbours, the chain decouples into to subchains, as can be seen by following the photon exchange loops above and below the chain.}
\label{fig:bnn_chainsplit}
\end{figure}

\begin{figure}
\begin{center}
\includegraphics{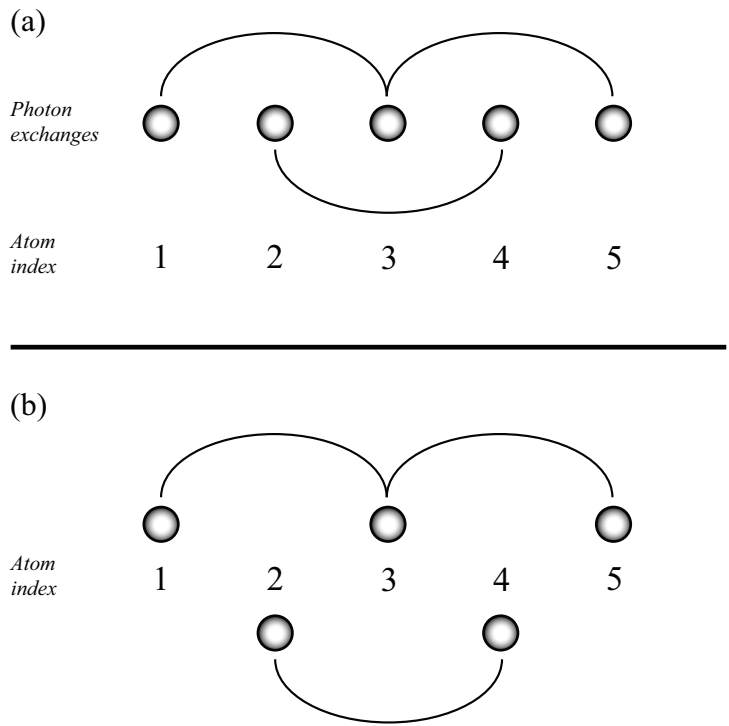}
\end{center}
\caption{(a) Next-nearest-neighbour interactions in a five atom chain.  (b)~The chain is visually separated into its two subchains; the fact that no photon lines connect them means that they do not interact with one another.}
\label{fig:bnn_chainsplit_odd}
\end{figure}

We now proceed to illustrate the explicit analytical solutions of (\ref{eq:TnFinalForm}) for some simplified cases of chains of coupled atoms. We first consider the case of no nearest-neighbour interaction, $a\rightarrow 0$. As illustrated in 
Figure \ref{fig:bnn_chainsplit}, turning off the nearest-neighbour interaction effectively reduces the chain into two distinct subchains that do not interact with each other.  Let us consider a concrete example, a chain of five atoms only, as shown in 
Figure \ref{fig:bnn_chainsplit_odd} (a).  Since the even-indexed atoms do not interact with the odd-indexed atoms, we are free to move them away from each other for clarity, as shown in Figure \ref{fig:bnn_chainsplit_odd} (b).  Notice that a photon that originated from either subchain has no way of being transferred to the other subchain. Thus, a five atom chain with the
nearest-neighbour interaction ignored is equivalent to two sub-chains, one with three and the other with two atoms.

This effect can also be seen in the Hamiltonian (\ref{eq:bnn_h}). When we put $a=0$ in the Hamiltonian, we obtain
\begin{equation}
H = \sum_{i=1}^{n} \omega_0 S_z^i + b \sum_{i \, \mathrm{even}} (S_i^+S_{i+2}^- + S_i^-S_{i+2}^+) + b \sum_{i \, \mathrm{odd}} (S_i^+S_{i+2}^- + S_i^-S_{i+2}^+) .
\end{equation}
It is evident that a photon in the ``even'' chain will never migrate to the ``odd'' chain, and vice versa.  

In practice, the procedure of removing the nearest-neighbour interaction can arise in two different ways:
\begin{itemize}
\item For critical values of $kr$ such that the nearest-neighbour coupling vanishes while the next-nearest-neighbour coupling does not.  This happens for example when $kr$ is equal to 4.48, 7.72, $\ldots$ for the case when $\boldsymbol{\mu} \bot \boldsymbol{r}_{ij}$.

\item Separating all the even-indexed atoms from the odd-indexed atoms by an infinite distance -- effectively forming two chains that are infinitely separated from each other.
\end{itemize}

The other case to consider is in the limit $b \rightarrow 0$, which switches off the next-nearest-neighbour interaction.  This is common practice when using the nearest-neighbour approximation.

\subsection{Rigorous solution in the limit of $a \rightarrow 0$}
\label{sec:soln_g_approaches_0}

We now illustrate the exact solution of the recurrence relation (\ref{eq:TnFinalForm}) for an arbitrary number of atoms with no nearest-neighbour interaction, $a\rightarrow 0$.  In this limit, the factors $\gamma e^{\pm i \alpha}$ that appear in the characteristic polynomial (\ref{eq:TnFinalForm}) reduce to
\begin{eqnarray}
\lim_{a \rightarrow 0} \gamma \rme^{\rmi \alpha} = \lim_{a \rightarrow 0} \left( \frac{\lambda + 2b + \sqrt{(\lambda + 2b)^2 - 4a^2}}{8b} \right)^{1/2} =  \sqrt{(\lambda + 2b)/4b},
\end{eqnarray}
and
\begin{eqnarray}
\lim_{a \rightarrow 0} \gamma \rme^{-\rmi \alpha} = \lim_{a \rightarrow 0} \left( \frac{\lambda + 2b - \sqrt{(\lambda + 2b)^2 - 4a^2}}{8b} \right)^{1/2} = 0 .
\end{eqnarray}
We also have
\begin{equation}
\lim_{a \rightarrow 0} 2a \sin 2 \alpha = \lim_{a \rightarrow 0} \sqrt{4a^2 - (\lambda + 2b)^2} = \rmi (\lambda + 2b).
\end{equation}
Therefore in the limit of $a \rightarrow 0$
\begin{eqnarray}
T_n = \frac{-b^{n+1}}{\lambda + 2b} \left\{ \left[ U_{n+1} \left( \sqrt{(\lambda + 2b)/4b} \right) \right]^2 - \left[ U_{n+1} \left( 0 \right) \right]^2 \right\} .\label{eq:Tn_lim_a0}
\end{eqnarray}
Equation (\ref{eq:Tn_lim_a0}) takes two different forms depending on whether $n+1$ is even or odd. If $n+1$ is an odd integer, then 
$U_{n+1} (0) = 0$, and we are left with the equation $U_{n+1} ( \sqrt{(\lambda + 2b)/4b} ) = 0$. 
This equation has $n+1$ roots given by $\sqrt{(\lambda + 2b)/4b} = \cos k \pi/(n+2)$, $1 \le k \le n+1$.  We disregard the trivial root  $\sqrt{(\lambda + 2b)/4b} = 0$ when $k = (n+2)/2$, since the trivial root is cancelled out by the factor $\lambda + 2b$ present in the denominator of $T_n$.  The remaining $n$ eigenvalues $E_k$ are given by
\begin{equation}
E_k = \omega_0 - 2b \cos \frac{2k \pi}{n+2},
\end{equation}
where $1 \le k \le n+1$, excluding $k = (n+2)/2$.  Since the eigenvalues associated with $k < (n+2)/2$ equal those with $k > (n+2)/2$, we can simplify the range of $k$ to $1 \le k \le n/2$, and state that the eigenvalues are \emph{doubly degenerate}.  This is consistent with the idea that the chain should behave like two chains of length $n/2$ in this case.

For $n+1$ even, $[U_{n+1} (0)]^2 = 1$, and therefore we look for the solutions of the equation 
$U_{n+1}( \sqrt{(\lambda + 2b)/4b} ) = \pm 1$, which are of the form
\begin{equation}
\sqrt{(\lambda + 2b)/4b} \in \left\{ \left. \cos \frac{k \pi}{n+1} \right| k \in I_1\right\}
\cup \left\{ \left. \cos \frac{k \pi}{n+3} \right| k \in I_2\right\} .
\end{equation}
This implies that the eigenvalues of the one-photon states are of the form
\begin{equation}
\lambda \in \left\{ \left. 2b \cos \frac{2k \pi}{n+1} \right| k \in I_1\right\}
\cup \left\{ \left. 2b \cos \frac{2k \pi}{n+3} \right| k \in I_2\right\} ,\label{e42}
\end{equation}
where $I_1$ and $I_2$ are sets of integers that we will determine below. 

The result (\ref{e42}) is the analytical solution for the energies of one-photon states of a chain of $n$ atoms with the next-nearest-neighbour interaction only. We see that the energies, and thus the energy states,  group into two separate manifolds. This is what we would expect from the above analysis  of a simplified problem of five atom chain that in the limit of $a=0$ the atom chain effectively decouples into two separate chains.  When the number of atoms is even, we have two sub-chains of equal length ($n/2$), and therefore doubly-degenerate energy levels.  When the chain has an odd number of atoms, it decouples into two chains of unequal length: $(n-1)/2$ and $(n+1)/2$.  This observation  leads us to the following restrictions on the values of $k$:
\begin{equation}
\fl \lambda \in \left\{ \left. 2b \cos \frac{2k \pi}{n+1} \right| k = 1, \ldots (n-1)/2\right\}
\cup \left\{ \left. 2b \cos \frac{2k \pi}{n+3} \right| k = 1, \ldots (n+1)/2\right\} ,
\end{equation}
which gives us a total of $n$ eigenvalues for the energies of the one-photon states.

\subsection{Rigorous solution in the limit of $b \rightarrow 0$}

We now illustrate the rigorous solution for the eigenvalues and eigenvectors in the other extreme case of no next-nearest-neighbour interaction, $b \rightarrow 0$.  According to (\ref{eq:gammadef}), taking the limit $b \rightarrow 0$ is equivalent to 
$\gamma \rightarrow \infty$.  

Expanding the Chebyshev polynomials in (\ref{eq:TnFinalForm}), we have
\begin{equation}
\fl T_n = \frac{-i b^{n+1}}{2a \sin 2 \alpha} \left\{ \left[ 2^{n+1} \gamma^{n+1} \rme^{\rmi (n+1) \alpha} + \cdots \right]^2 - \left[ 2^{n+1} \gamma^{n+1} \rme^{-\rmi (n+1) \alpha} + \cdots \right]^2 \right\}.
\end{equation}
Clearly, the terms with the highest power of $\gamma$ will dominate the expression.  We therefore have in the limit of large $\gamma$
\begin{equation}
T_n = \frac{-i}{2a \sin 2 \alpha} \left[ 2^{2(n+1)} b^{n+1} \gamma^{2(n+1)} \left( \rme^{\rmi 2(n+1) \alpha} - \rme^{-\rmi 2(n+1) \alpha} \right) \right].
\end{equation}
Since $b^{n+1} \gamma^{2(n+1)} = (a/4)^{n+1}$, we finally arrive at
\begin{equation}
\lim_{\gamma \rightarrow \infty} T_n = a^{n} \frac{\sin (n+1) 2 \alpha}{\sin 2 \alpha}.
\end{equation}
The above function clearly has $n$ roots (since it has a Chebyshev polynomial of the second kind).  These solutions are given by $2 \alpha_k = k \pi / (n+1)$ for $1 \le k \le n$.  We thus obtain the energy eigenvalues
\begin{equation}
E_k = \lim_{b \rightarrow 0} \left( \omega_0 + 2b - 2a \cos 2\alpha_k \right) = \omega_0 - 2a \cos \frac{k \pi}{n+1},
\end{equation}
which agrees with the result of Lieb, Schultz and Mattis \cite{bib:Lieb1961}, who made use of the Jordan-Wigner transform to diagonalize the Hamiltonian of the nearest-neighbour $XY$ chain.  An alternative derivation that does not use the Jordan-Wigner transform, but is restricted to the $XX$ chain, is given in \cite{bib:Mewton2005}.

\section{Power series solution of the general characteristic polynomial}

In this section, we will outline a power series approximation of finding the roots of the general characteristic polynomial (\ref{eq:TnFinalForm}).  In order to find the roots of $T_n$, let us look closely into the procedure of solving the following equation
\begin{equation}
\left[ U_{n+1} \left( \gamma \rme^{\rmi \alpha} \right) \right]^2
-  \left[ U_{n+1} \left( \gamma \rme^{-\rmi \alpha} \right) \right]^2 = 0 .
\end{equation}
Equivalently, this equation may be written in terms of two separate equations
\begin{equation}
U_{n+1} \left( \gamma \rme^{\rmi \alpha} \right)
-  U_{n+1} \left( \gamma \rme^{-\rmi \alpha} \right) = 0,
\label{eq:Umin}
\end{equation}
and
\begin{equation}
U_{n+1} \left( \gamma \rme^{\rmi \alpha} \right)
+  U_{n+1} \left( \gamma \rme^{-\rmi \alpha} \right) = 0,
\label{eq:Uplus}
\end{equation}

In general, the solutions $\alpha$ will either be real or imaginary, depending on the value of $\gamma$.  The aim of this section is to develop a power series solution, so it would be helpful to work with some property of $\alpha$ which does not rotate in the complex plane as $\gamma$ varies.  Let us define
\begin{eqnarray}
x & \equiv & \smallfrac{1}{2} \left( \arccos \gamma \rme^{\rmi \alpha} + \arccos \gamma \rme^{-\rmi \alpha} \right), \label{eq:xdef} \\
y & \equiv & \smallfrac{1}{2} \left( \arccos \gamma \rme^{\rmi \alpha} - \arccos \gamma \rme^{-\rmi \alpha} \right).
\end{eqnarray}

Since we know that $\alpha$ must be purely real or imaginary, it is easy to see that $x$ as defined above will always be real, just the property we were looking for.  It is the goal of this section to, therefore, develop power series solutions of $x$ for (\ref{eq:Umin}) and (\ref{eq:Uplus}).  To do this, it is helpful to express (\ref{eq:Umin}) and (\ref{eq:Uplus}) in a different form.  Let us consider the following equation
\begin{equation}
\frac{\sin (n+2)(x+y)}{\sin (x+y)} \pm \frac{\sin (n+2)(x-y)}{\sin (x-y)} = 0,
\end{equation}
where we obtain (\ref{eq:Umin}) when the $\pm$ operation is replaced by a minus sign, and the converse for (\ref{eq:Uplus}).  

Expanding out the trigonometric functions to separate $x$ and $y$, we obtain
\begin{eqnarray}
&&[ \sin x \cos y - l \cos x \sin y ] \nonumber \\
&&\times [ \sin (n+2) x \cos (n+2) y + l \cos (n+2) x \sin (n+2) y ] \pm \mbox{conj.}  =  0.
\end{eqnarray}
We have introduced the parameter $l$ into the above expression, which is set equal to unity.  The ``conj.'' term is merely the first term in the expression but with the $l$'s set equal to $-1$.  Expanding away the square brackets gives
\begin{eqnarray}
&& \sin x \cos y \sin (n+2) x \cos (n+2) y \nonumber \\
&& - l^2 \cos x \sin y \cos (n+2) x \sin (n+2) y \nonumber \\
&& - l \cos x \sin y \sin (n+2) x \cos (n+2) y \nonumber \\
&& + l \sin x \cos y \cos (n+2) x \sin (n+2) y \pm \mbox{conj.} = 0.
\label{eq:TangBoth}
\end{eqnarray}
We are now presented with two expressions, corresponding to the choice made with the $\pm$ operator.  Taking the negative case causes all terms with an even power of $l$ (or zero) to cancel each other out, leaving
\begin{eqnarray}
&& \cos x \sin y \sin (n+2) x \cos (n+2) y = \sin x \cos y \cos (n+2) \sin (n+2) y \nonumber \\
&& \Longleftrightarrow \tan y \tan (n+2) x = \tan x \tan (n+2) y.
\label{eq:Tmin}
\end{eqnarray}
This expression is equivalent to (\ref{eq:Umin}).
We now consider the positive branch of the $\pm$ operator in (\ref{eq:TangBoth}).  In this case, the terms proportional to an odd power of $l$ cancel each other out, leaving
\begin{eqnarray}
&& \sin x \cos y \sin (n+2) x \cos (n+2) y = \cos x \sin y \cos (n+2) x \sin (n+2) y \nonumber \\
&& \Longleftrightarrow \tan x \tan (n+2) x = \tan y \tan (n+2) y.
\label{eq:Tplus}
\end{eqnarray}

We now look for power series solutions.  For simplicity, we will assume that the chain is of even length.  However, the following method will in principle work for chains of odd length, although the resulting power series will be more complicated.  Consider first (\ref{eq:Tmin}).  Using the relation
\begin{equation}
y = \smallfrac{1}{2} \arccos ( 2 \gamma^2 - \cos 2x ),
\end{equation}
we can express $y$ as a function of $x$, and we let $x$ to be represented by the power series
\begin{equation}
x = \sum_{k=0}^{\infty} x_k \gamma^k.
\label{eq:PowerMin}
\end{equation}

We next expand both sides of (\ref{eq:Tmin}) in powers of $\gamma$.  Since we want the power series solution (\ref{eq:PowerMin}) to be valid for all $\gamma$, we require that all the coefficients of all the powers of $\gamma$ vanish.  Doing this gives us the following power series:
\begin{eqnarray}
x & = & x_0 - \frac{\sin x_0}{2 \cos x_0} \gamma^2 - \frac{2 \cos^2 x_0 - 1}{8 \sin x_0 \cos^3 x_0} \gamma^4 \nonumber \\
&& + \left[ 4 (n+2)^2 \cos^8 x_0 + 6 \cos^6 x_0 - 6 (n+2)^2 \cos^6 x_0 \right. \nonumber \\
&&\left. - 8 \cos^4 x_0 + 2 (n+2)^2 \cos^4 x_0 + 8 \cos^2 x_0 -3 \right] \nonumber \\
&& \times (48 \sin^3 x_0 \cos^5 x_0)^{-1} \gamma^6  + \cdots ,
\end{eqnarray}
where $x_0$ is one of the solutions for the case $\gamma = 0$.  For $n$ even, we have
\begin{equation}
x_0 = \pi/2 + 2k \pi / (n+2), \quad 1 \le k \le n/2.
\label{eq:PowerSolnX0}
\end{equation}

A power series can be similarly obtained for (\ref{eq:Tplus}).  In this case, we find that

\begin{eqnarray}
x & = & x_0 - \frac{\cos x_0}{2 \sin x_0} \gamma^2 - \frac{2 \cos^2 x_0}{8 \cos x_0 \sin^3 x_0} \gamma^4 \nonumber \\
&& + \left[ 4 (n+2)^2 \cos^8 x_0 - 6 \cos^6 x_0 - 10 (n+2)^2 \cos^6 x_0 \right. \nonumber \\
&& \left. + 10 \cos^4 x_0 + 8 (n+2)^2 \cos^4 x_0 -10 \cos^2 x_0 - 2 (n+2)^2 \cos^2 x_0 + 3 \right] \nonumber \\
&& \times (48 \sin^5 x_0 \cos^3 x_0)^{-1} \gamma^6 + \cdots .
\end{eqnarray}
For this case, $x_0$ is simply one of the solutions to given in (\ref{eq:PowerSolnX0}).

Given any power series, one should consider the radius of convergence.  Since we do not have a tractable formula for $[\gamma^k]x(\gamma)$ for arbitrary $k$, we cannot investigate convergence properties.  The power series presented here, however, have another important purpose.  They provide a way to work out all the derivatives of $x(\gamma)$.  This would be useful if a way could be found to formulate a differential equation to solve to give a closed form for $x(\gamma)$.  The terms of the power series would then provide as many boundary conditions as needed at $\gamma = 0$.  It may also be possible to work out the generating function for the series.  Some of the terms can be summed over all powers of $\gamma$, but there are many other terms which are too intractable.  Again, it is important to emphasize that power series terms given in this section are for chains of even length.

\section{General properties}

In this section, we shall discuss various properties of the function $x(\gamma)$ given by~(\ref{eq:xdef}).  We shall not limit ourselves in this section to a power series solution.  All results were obtained by solving (\ref{eq:TnFinalForm}) numerically for fixed values of $n$.

\begin{figure}
\begin{center}
\includegraphics{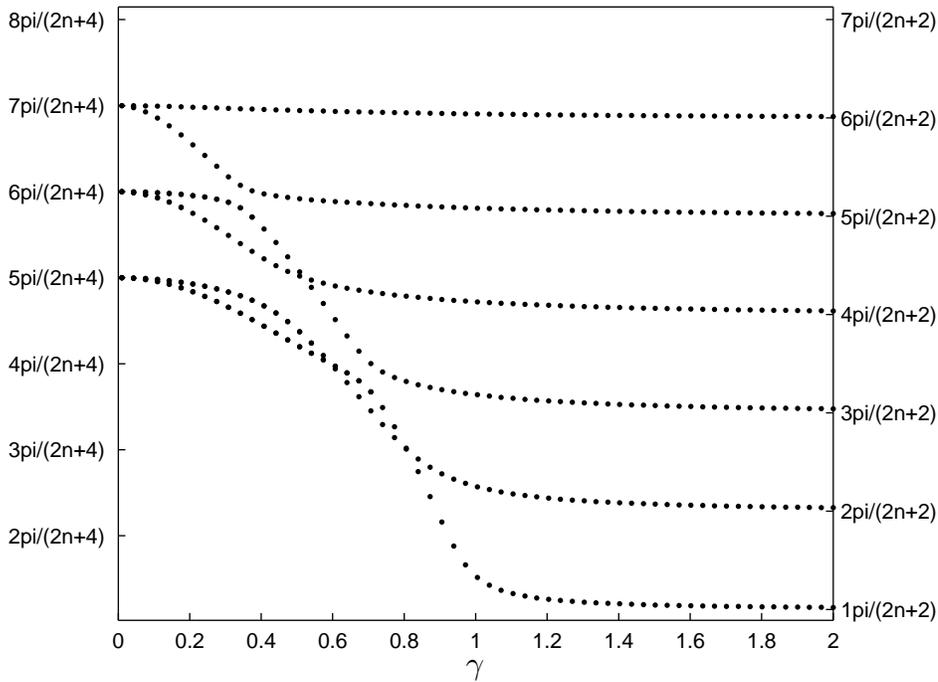}
\end{center}
\caption{The function $x(\gamma)$ plotted for a 6-atom chain.}
\label{fig:bnn:n6xplot}
\end{figure}

Working with the function $x$ offers the convenience of investigating the energy levels as $\gamma$ varies from zero to infinity.  The function $x(\gamma)$ always remains finite, as can be seen in Figure \ref{fig:bnn:n6xplot}.  Each curve in the graph is connected with a particular energy eigenvalue of the system.  This graph typifies the behaviour of chains with an even number of atoms.  The graph shows which states corresponding to $\gamma=0$ ($a=0$, next-nearest-neighbour-only interaction)  transition into the states associated with $\gamma \rightarrow \infty$ ($b=0$, nearest-neighbour-only interaction).  To show the relationship between $x(\gamma)$ and the energy eigenvalues, we can briefly calculate $x(\gamma)$ analytically for the cases $\gamma = 0$ and $\gamma \rightarrow \infty$.  In the former case, we have (for chains of even length) that
\begin{eqnarray}
x(0) &=& \smallfrac{1}{2} \left( \arccos \lim_{\gamma \rightarrow 0} \gamma \rme^{\rmi \alpha} + \arccos \lim_{\gamma \rightarrow 0} \gamma \rme^{-\rmi \alpha} \right)\nonumber \\
&=& \smallfrac{1}{2} \arccos \sqrt{ (\lambda + 2b)/4b} + \smallfrac{1}{2} \arccos 0 =  \frac{k\pi}{2(n+2)} + \frac{\pi}{4},
\end{eqnarray}
where we have used the results presented in Sec.~\ref{sec:soln_g_approaches_0} to obtain the above result, and $1 \le k \le n/2$.  Therefore, we expect a total of $n$ solution curves of $x(\gamma)$ to intersect the $\gamma=0$ axis at points $k \pi / 2(n+2) + \pi / 4$, $1 \le k \le n/2$, with two curves intersecting the axis at these specific points.  A cursory glance at Figure \ref{fig:bnn:n6xplot} shows this to be the case.

At the other extreme, we can consider the values of $x(\gamma)$ in the limit $\gamma \rightarrow \infty$.  In this limit, we know that $\alpha \rightarrow \alpha_k \equiv k \pi /2(n+1)$, for $1 \le k \le n$.  We therefore have:
\begin{eqnarray}
\lim_{\gamma \rightarrow \infty} x(\gamma) &=& \smallfrac{1}{2} \left( \arccos \lim_{\gamma \rightarrow \infty} \gamma \rme^{\rmi \alpha} + \arccos \lim_{\gamma \rightarrow \infty} \gamma \rme^{-\rmi \alpha} \right)\nonumber \\
&=& \smallfrac{1}{2} \lim_{\gamma \rightarrow \infty} (\arccos \gamma \rme^{\rmi \alpha_k} +  \arccos \gamma \rme^{-\rmi \alpha_k}) \nonumber \\
&=& \smallfrac{1}{2} \lim_{\gamma \rightarrow \infty} \arccos \left( \gamma^2 - \sqrt{1-\gamma^2 \rme^{\rmi 2 \alpha_k}} \sqrt{1-\gamma^2 \rme^{-\rmi 2 \alpha_k}} \right) \nonumber \\
&=& \smallfrac{1}{2} \arccos \cos 2 \alpha_k = \alpha_k .
\end{eqnarray}
Indeed, the solution curves in Figure \ref{fig:bnn:n6xplot} for $x(\gamma)$ approach the $\alpha_k$ in the limit of large $\gamma$; for convenience, the right-hand vertical axis is marked off in terms $\alpha_k \equiv k \pi /2(n+1)$ to clearly show this.  In the case of our specific graph, $\gamma=2$ is large enough to show this limiting behaviour.

Where the lines cross, the eigenstates associated with the two levels are clearly degenerate.  We therefore have the interesting effect that including the next-nearest neighbour effect can induce degeneracy, which is not present in one-photon states when $\gamma \rightarrow \infty$ (i.e. $b=0$).  These states are degenerate, however, when $\gamma = 0$.  When considering these pairs of degenerate states, we see that the highest pair has only one degenerate point (at $\gamma = 0$).  The second pair underneath has degeneracies for two values of $\gamma$, while the third pair has degeneracies occurring at three values of $\gamma$.  This is a general trend for chains of even length.

\begin{figure}
\begin{center}
\includegraphics{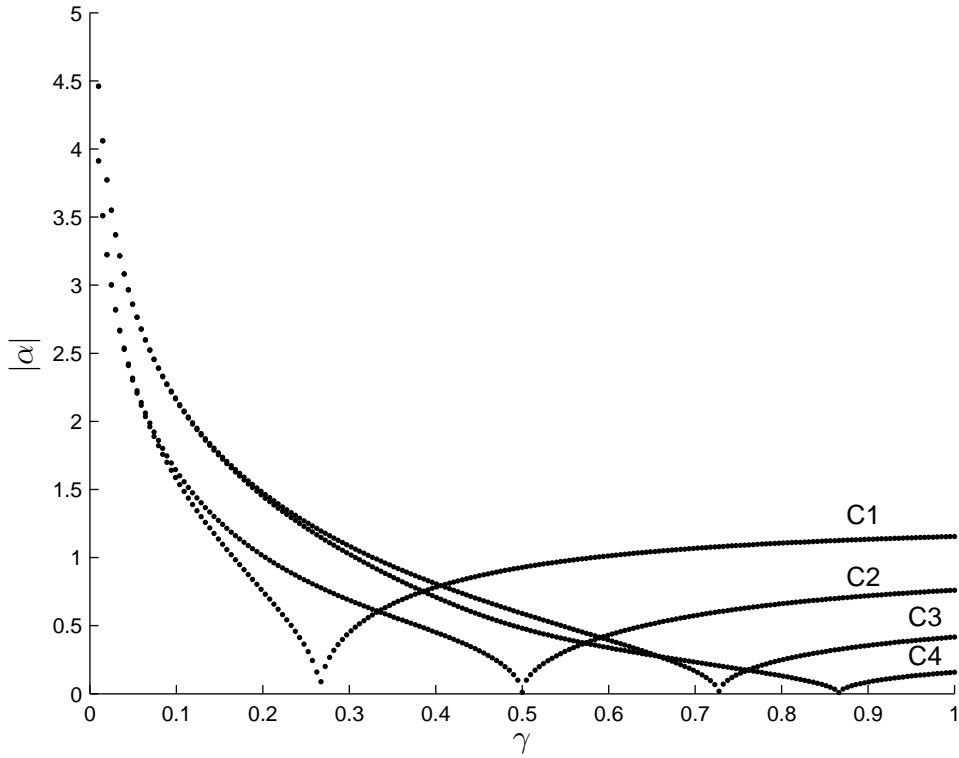}
\end{center}
\caption{The multi-valued function $|\alpha(\gamma)|$ plotted for a 4-atom chain.  There are four solution curves, $C1$, $C2$, $C3$ and $C4$.}
\label{fig:bnn:n4_abs_alpha}
\end{figure}

\begin{figure}
\begin{center}
\includegraphics{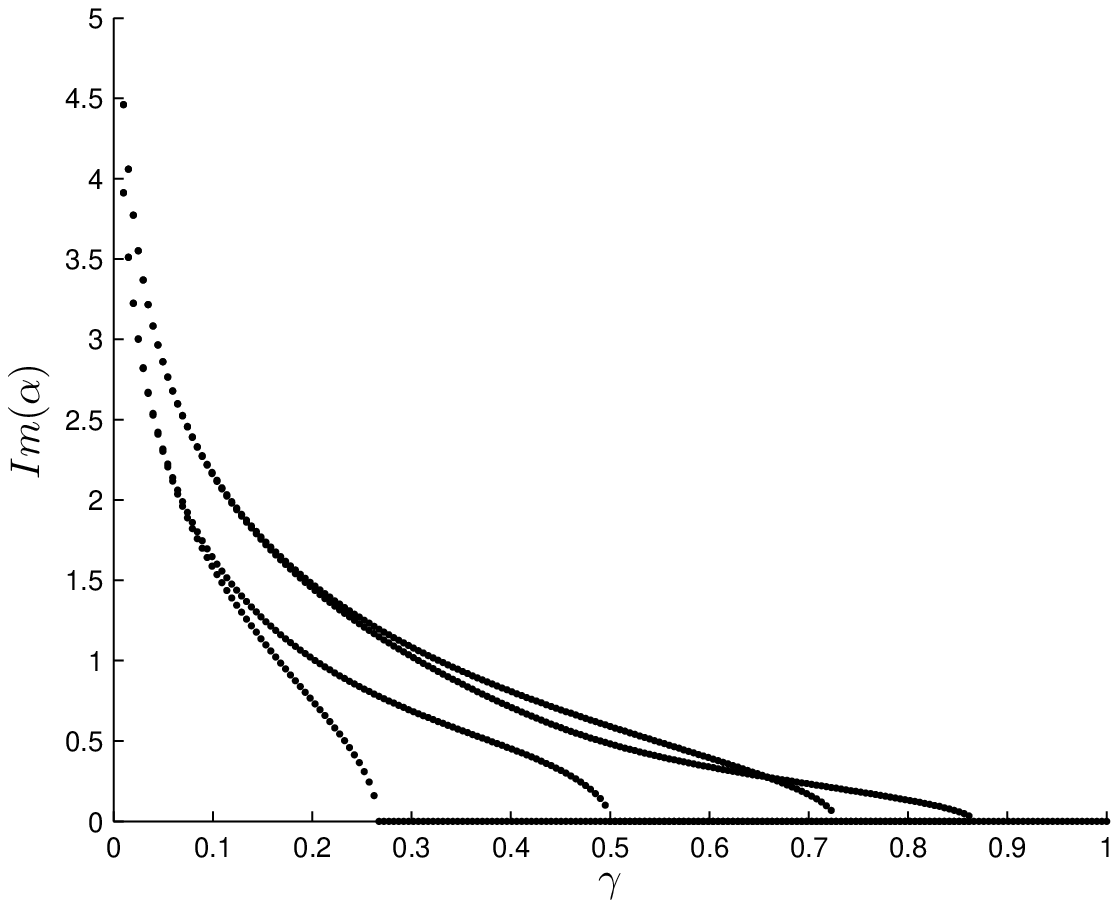}
\end{center}
\caption{The imaginary part of $\alpha(\gamma)$ plotted for a 4-atom chain for the four solution curves.}
\label{fig:bnn:n4_imag_alpha}
\end{figure}

The solutions in terms of $\alpha$ also have interesting properties.  These are obtained by simply solving (\ref{eq:TnFinalForm}) for $\alpha$, which has to be done numerically for the time being, since an analytic solution has not been found yet for arbitrary $\gamma$.  Consider Figure \ref{fig:bnn:n4_abs_alpha}, which shows the absolute value of the solutions $\alpha$ for a 4-atom chain.  We see that each solution curve $C$ touches the $\gamma$ axis at only one location, $\gamma_0(C)$.  Furthermore, the curve to the left of $\gamma_0(C)$ is purely imaginary; to the right, it is real.  Figure \ref{fig:bnn:n4_imag_alpha} shows only the imaginary component of $\alpha$; the plot clearly supports this conclusion.  This is a general trend for positive $\gamma$.  Based on similar graphs for longer chains, these real-imaginary crossover points always occur for $\gamma < 1$; we can therefore conclude that for $\gamma \ge 1$, all solutions are real for positive $\gamma$.  It is our hope that by pointing out this general feature that the solution curves turn through a right angle in the complex plane for some critical value of $\gamma$ for $\gamma < 1$, light is shed on the possible forms an analytic solution could have.

\section{Energy eigenstates}

Given the eigenvalues $E_i$ associated with the Hamiltonian $H$ given by (\ref{eq:bnn_h}), the corresponding eigenvectors $\ket{\psi_i}$ for single-photon excitations can be derived as follows.  Let the eigenvector $\ket{\psi_i}$ be given by a linear superposition 
\begin{equation}
\ket{\psi_i} = \sum_{j=1}^n c_j \ket{j},
\end{equation}
where $\ket{j}$ is an eigenvector of $S_k^z$ for all $1 \le k \le n$.  The eigenvalue equation $H \ket{\psi_i} = E_i \ket{\psi_i}$ implies that
\begin{equation}
b( c_{j-2} + c_{j+2} ) + a( c_{j-1} + c_{j+1} ) = E_i c_j,
\label{eq:eigvecRecurrence}
\end{equation}
for $2 < j < (n-1)$ (and we have set $\omega_0=0$ without loss of generality for simplicity), and
\begin{eqnarray}
&& ac_2 + bc_3 = E_i c_1, \label{eq:eigvecRecurrenceA}\\
&& a(c_1 + c_3) + bc_4 = E_i c_2, \label{eq:eigvecRecurrenceB}\\
&& a(c_{n-2} + c_n ) + bc_{n-3} = E_i c_{n-1}, \label{eq:eigvecRecurrenceC}\\
&& ac_{n-1} + bc_{n-2} = E_i c_n. \label{eq:eigvecRecurrenceD}
\end{eqnarray}
Equation (\ref{eq:eigvecRecurrence}) is valid for all $1 \le j \le n$ provided we impose the boundary conditions
\begin{equation}
c_{-1} = c_0 = c_{n+1} = c_{n+2} = 0.
\label{eq:eigvecBoundaryConditions}
\end{equation}
These boundary conditions merely force (\ref{eq:eigvecRecurrence}) to reproduce (\ref{eq:eigvecRecurrenceA}), (\ref{eq:eigvecRecurrenceB}), and (\ref{eq:eigvecRecurrenceC}).  We now solve the recurrence relation (\ref{eq:eigvecRecurrence}).  Let $c_j = x^j$. 
Then (\ref{eq:eigvecRecurrence}) becomes
\begin{eqnarray}
b(x^{j-2} + x^{j+2}) + a(x^{j-1}+x^{j+1}) & = & E_i x^j \nonumber \\
\Longleftrightarrow 2b(2 \cosh^2 \ln x - 1) + 2a \cosh \ln x & = & E_i .
\end{eqnarray}
Let $z = 2 \cosh \ln x$.  Then
\begin{equation}
b(z^2 - 2) + az = E_i.
\end{equation}
This quadratic expression has the two solutions
\begin{equation}
z_{\pm} = \frac{ -a \pm \sqrt{a^2 + 4b(2b+E_i)} }{2b}.
\end{equation}
We then have two solutions for $x$:
\begin{equation}
x_{1} = e^{\pm \arccosh (z_+/2)} \quad \mbox{and} \quad x_{2}=e^{\pm \arccosh (z_-/2)}.
\end{equation}
We can therefore write the coefficients $c_j$ as
\begin{eqnarray}
c_j &=& A e^{j \arccosh (z_+/2)} + B e^{-j \arccosh (z_+/2)} \nonumber \\
&& + C e^{j \arccosh (z_-/2)} + D e^{-j \arccosh (z_-/2)},
\end{eqnarray}
where $A$, $B$, $C$, and $D$ are arbitrary constants whose values are derived by imposing the boundary conditions (\ref{eq:eigvecBoundaryConditions}).  We notice, however, that we have four boundary conditions, thus the naive solution to this system of equations is that $A$, $B$, $C$, and $D$ vanish.  However, these four boundary conditions cannot be linearly independent; one of the arbitrary constants should be undetermined so that we can impose an arbitrary normalisation on the eigenstate.  This means that we really have three independent equations.  Without knowing the specific form of the energy eigenvalue as a function of $a$ and $b$, however, we cannot tell which boundary condition is a trivial consequence of the other three.  Therefore, we stop the derivation at this point.  We have laid out most of the calculation so that once the energy eigenvalues can be completely determined, the remaining steps to determine the eigenstates are simple.

\section{Conclusion}

We have derived the characteristic polynomial of the eigenvalue equation for a chain of atoms with up to the next nearest-neighbour coupling. The exact form of the polynomial is given in terms of the Chebyshev polynomials of the second kind and is valid for an arbitrary number of atoms and coupling strengths. We have illustrated the exact analytical solutions of the characteristic polynomial for two simplified cases of no nearest-neighbour coupling and no next-nearest-neighbour coupling.

We have also derived the power series solution of the general characteristic polynomial. We have shown that there are patterns as to how many times a pair of eigenstates become degenerate. 
We have also outlined a simple approach by which the energy eigenstates for single excitations may be found.

We hope that our presentation provides tools for further advancement towards an analytic solution for the energy eigenvalues of arbitrary excited linear chains and
in the analytic studies and understanding of the energy structure of linear chains of coupled atoms  and also to simplification of the mathematics involved. This alone hopefully is advantageous since it helps reduce the number of operations a computer has to perform to find the energy eigenvalues.

\section*{References}


\providecommand{\newblock}{}

\end{document}